# Gravitational Waves III: Detecting Systems

M. Cattani
Instituto de Fisica, Universidade de S. Paulo, C.P. 66318, CEP 05315-970
S. Paulo, S.P. Brazil . E−mail: mcattani@if.usp.br

Abstract.
In a recent paper we have deduced the basic equations that predict the emission of gravitational waves (GW) according to the Einstein gravitation theory. In a subsequent paper these equations have been used to calculate the luminosities and the amplitudes of the waves generated by binary stars, pulsations of neutron stars, wobbling of deformed neutron stars, oscillating quadrupoles, rotating bars and collapsing and bouncing cores of supernovas. We show here how the GW could be detected in our laboratories. This paper, like the preceding ones, was written to graduate and postgraduate students of Physics.

Resumo.
Num artigo recente deduzimos as equações básicas que prevêem a emissão de ondas gravitacionais (OG) de acordo com a teoria de gravitação de Einstein. Num artigo posterior usamos essas equações para estimar as luminosidades e amplitudes de ondas geradas por sistemas emissores tais como estrelas binárias, pulsações de estrelas de nêutrons, precessão de estrelas de nêutrons deformadas, quadrupolos oscilantes, barras girantes e nos processos cataclísmicos que dão origem a supernovas. Veremos agora como as OG poderiam ser detectadas em nossos laboratórios. Esse trabalho, tal como os anteriores, foi escrito para alunos de graduação e pós−graduação em Física.

## I. Introdução

Todas as teorias relativísticas aceitas sobre gravitação [1-4] prevêem a existência de ondas gravitacionais (OG). Num artigo recente,[5] assumindo a teoria de gravitação de Einstein como sendo a mais fidedigna, deduzimos as equações gerais que prevêem a emissão de ondas gravitacionais (OG). No artigo seguinte[6] usamos essas equações para estimar as luminosidades e amplitudes de ondas geradas pelos seguintes sistemas emissores: estrelas binárias, pulsações de estrelas de nêutrons, precessão de estrelas de nêutrons deformadas, quadrupolos oscilantes, barras girantes e nos processos cataclísmicos que dão origem a supernovas. Agora mostraremos como detectar as OG que em princípio se baseiam na medida de deslocamentos relativos de massas e de suas variações com o tempo. Conforme veremos nas próximas seções, existem muitos sistemas detectores diferentes que são usados e outros estão sendo projetados:[3,7,8] :
interferométricos, sólidos ressonantes, oscilações da distância entre Terra e
Lua, oscilações da crosta da Terra, modos normais de vibração de sólidos
em forma de retângulos, forquilhas e argolas, barras girantes e tubos
girantes com fluidos dentro. Atualmente, os mais cotados, ou seja, os que
parecem ter mais chance de detectar as OG são os *Interferométricos* e os
*Sólidos Ressonantes*. Estes são constituídos por sólidos muito rígidos, em
geral, em forma de barras cilíndricas ou esferas.[3,7,8]. Somente esses dois
tipos serão analisados por nós.

**1. Detectores Interferométricos**.

O detector interferométrico é basicamente um grande interferômetro
de Michelson[7-9] a laser, com braços perpendiculares, onde há três espelhos
M, $M_1$ e $M_2$ conforme Figura 1. Os espelhos M1 e M2 são presos a blocos
suspensos que podem oscilar livremente como pêndulos. Com o auxílio de
um laser altamente monocromático e muito potente o interferômetro mede
o deslocamento relativo dos espelhos que seria gerado pela OG.

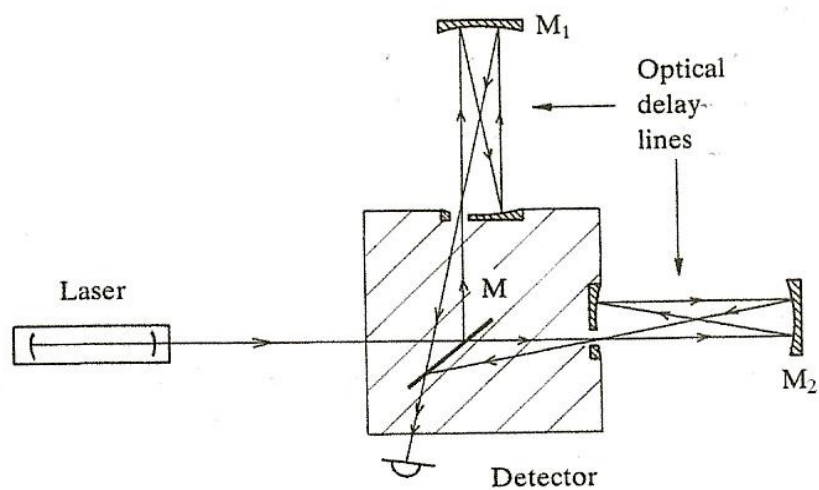

Figura 1. Esquema de um típico grande interferômetro de Michelson para
ser usado na detecção de ondas gravitacionais.[7]

Atualmente vários detectores interferométricos estão sendo
construídos e projetados: GEO 600 (colaboração entre Alemanha e
Inglaterra), LIGO (USA), VIRGO (colaboração entre Itália e França),
TAMA 300 (Japão), AIGO (Austrália), LISA (projeto espacial, NASA,
ESA) e LISC (LISA− projeto espacial internacional). Como há uma imensa
quantidade de informações técnicas envolvidas nesses grandes projetos
sugerimos aos leitores que entrem no Google usando as palavras
"Interferometric Detectors".

Analisemos o princípio de funcionamento desses detectores. Assim,
(vide Fig.1) assumiremos que a OG incida na direção do eixo z

(perpendicular ao plano da figura) com os eixos de polarização (+) ao longo dos eixos x (passando pelo laser, M e $M_2$) e y (passando por M e $M_1$). Seguindo o nosso primeiro artigo,[5] vamos ver como essa OG muda a distância entre duas partículas localizadas ao longo do eixo x (y = z =0), uma no ponto x = −dx/2 e a outra em x = dx/2. Supondo que a OG tenha uma única polarização (+) com amplitude h, a distância dℓ(x,t) (calculada para tempos retardados t − x/c) entre as partículas de acordo com a Eq.(III.3.2) vista no Apêndice III,[5] é dada por

$$dℓ(x,t) ≈ [1−(h/2)\cos(ωt − kx)] \, dx \qquad (1.1),$$

onde k = ω/c = 2π/λ sendo ω e λ, respectivamente, a freqüência angular e o comprimento de onda da OG. Assim, se os pontos estiverem separados por uma distância finita com coordenadas x = ±$ℓ_o$/2 de (1.1) obtemos a *distância* ℓ(t)

$$ℓ(t) = \int_{-ℓ_o/2}^{ℓ_o/2} [1−(h/2)\cos(ωt − kx)]dx = ℓ_o + (h/2) \int_{-ℓ_o/2}^{ℓ_o/2} \cos(ωt − kx) \, dx \qquad (1.2),$$

onde $ℓ_o$ é a distância antes da OG ter chegado. Assim, de acordo com (1.2) a alteração da distância Δℓ(t) = ℓ(t) − $ℓ_o$ gerada pelas OG é dada por

$$Δℓ(t) = (h/k) \cos(ωt) \sin(kℓ_o/2) = (λh/2π) \cos(ωt) \sin(πℓ_o/λ) \qquad (1.3),$$

mostrando que a **amplitude** Δℓ do deslocamento é dada por

$$Δℓ = (λh/2π) \sin(πℓ_o/λ) \qquad (1.4),$$

Em condições análogas para dois pontos ao longo do eixo y o Δℓ é dado também por (1.4). Notemos que para $ℓ_o$→ 0 a amplitude de deformação ε = Δℓ/$ℓ_o$ → h/2 que é a amplitude do cisalhamento[5] ao longo do eixo x ou y, como era de se esperar. As variações de distância Δℓ desses braços têm fases contrárias: enquanto um encurta o outro alonga e vice−versa.[5]

De acordo com (1.4) o valor máximo de Δℓ é obtido quando $ℓ_o$= λ/2, ou seja, (Δℓ)$_{max}$ = h λ/2π. Isto implica que a amplitude mínima $h_{min}$ de uma OG capaz de provocar uma alteração de distância Δℓ é dada por $h_{min}$ ≈ Δℓ/(λ/2). As mudanças de distâncias Δℓ(t) (vide (1.3)) causam uma correspondente oscilação nas franjas de interferência. A luz laser é dividida em duas partes pelo espelho M que se propagar ao longo dos braços do interferômetro. Os feixes são refletidos por M1 e M2, retornam, recombinam coerentemente em M novamente e são detectados por dispositivo fotossensível. Na (1.4) a distância $ℓ_o$ seria o *"comprimento do caminho ótico"* não perturbado. De (1.4) vemos que o valor máximo de Δℓ é obtido quando $\ell_o = \lambda/2$, ou seja, $(\Delta\ell)_{max} = \ell_o h/\pi$. Para uma OG com f = 1kHz, $(\Delta\ell)_{max}$ é obtido quando $\ell_o$= 150 km. Múltiplas passagens em cada braço são efetuadas para aumentar o caminho ótico obtendo desse modo um "*comprimento ótico efetivo*" L. Um esquema simplificado é mostrado na Fig.1 com os feixes indo e voltando entre os espelhos por sucessivas reflexões. Esse processo é conhecido como "*linha de atraso*"(optical delay line). Supondo que h = $10^{-21}$ e f = 1 kHz, ou seja, que $\lambda$ = 300 km verificamos que $(\Delta\ell)_{max} = h \lambda/2\pi = 5\ 10^{-15}$ cm, isto é, 100 vezes menor do que o raio nuclear. No observatório VIRGO, por exemplo, proposto em 1990[10] e que está sendo atualmente construído em Cascina, próximo a Pisa (Itália), os braços têm 3 km de extensão e espera−se ter L = 120 km.

Devido às pequeníssimas intensidades das OG[5,6] o interferômetro deverá ser oticamente perfeito e extremamente bem isolado do resto do mundo (isolamento sísmico, de raios cósmicos, campos eletromagnéticos, etc.). Os lasers devem ser muito potentes, extremamente monocromáticos e estáveis, os espelhos devem ter altíssima refletividade, o percurso da luz deve ser feito ao longo de tubos com extremo alto vácuo, etc.

A precisão das medidas está fundamentalmente limitada pelas flutuações geradas pelas flutuações, nas franjas de interferência, do número de fótons detectados, pois elas simulam o efeito de mudanças do caminho ótico. Suponhamos que no ponto P (no espelho M) a intensidade da luz proveniente de um braço seja A e que a intensidade proveniente do outro braço seja $A \exp(i2\pi\Delta L/\lambda_{em})$, onde $\Delta L = L_1 - L_2$ é diferença dos *caminhos efetivos* dos dois feixes de luz ao longo dos braços 1 e 2, respectivamente, e $\lambda_{em}$ é o comprimento de onda da luz laser. Assim, a intensidade da luz no detector é dada por

$$| A + A \exp(i2\pi\Delta L/\lambda_{em}) |^2 = 4 A^2 \cos^2(\pi\Delta L/\lambda_{em}) \qquad (1.5)$$

A distribuição do número de fótons N nas franjas de interferência é, então, dada por[9]

$$N = N_{max} \cos^2(\pi\Delta L/\lambda_{em}) \qquad (1.6).$$

Devido a OG incidente as distâncias originalmente não perturbadas $L_1$ e $L_2$ são alteradas de $dL_1$ e $dL_2$ resultando numa variação total $d(\Delta L) = dL_1 - dL_2$. O efeito de $d(\Delta L)$ sobre N é dado por

$$dN = N_{max} (\pi/\lambda_{em}) \sin(2\pi\Delta L/\lambda_{em}) d(\Delta L) \qquad (1.7).$$

Como o erro estatístico de N é $N^{1/2}$ a precisão na medida de $d(\Delta L)$ pode ser estimada substituindo dN por $N^{1/2}$ em (1.7) obtendo

$$d(\Delta L) = \lambda_{em} N^{1/2}/[\pi N_{max} \sin(2\pi\Delta L/\lambda_{em})] \qquad (1.8),$$

de onde verificamos que d(ΔL) é mínimo quando $\sin(2\pi\Delta L/\lambda_{em})=1$. Com essa condição a (1.6) fica escrita como $N = N_{max}/2 = N_o$, onde $N_o$ representa a intensidade média. Nessas condições, de (1.8) decorre

$$d(\Delta L) = \lambda_{em}/(2\pi N_o^{1/2}) \qquad (1.9).$$

O valor de $N_o$ depende da potência P do laser, do intervalo de tempo $\Delta\tau$ de duração da medida e da eficiência ε de detecção dos fótons de acordo com a equação[7,9]

$$N_o = \Delta\tau \, \varepsilon \, P/(\hbar\omega_{em}) \qquad (1.10),$$

onde $\omega_{em}$ é a freqüência angular da luz laser. Substituindo (1.10) em (1.9) temos

$$d(\Delta L) = (\lambda_{em}/2\pi)(\hbar\omega_{em}/\Delta\tau \, \varepsilon \, P)^{1/2} = [c^2\hbar/\omega_{em}\Delta\tau \, \varepsilon \, P]^{1/2} \qquad (1.11)$$

Nós vimos usando (1.4) que o valor máximo de deslocamento devido a OG é obtido quando o comprimento do percurso era igual a λ/2. Nessas condições os fótons permanecem um tempo (de percurso) λ/2c dentro do braço do interferômetro. Isto significa que o tempo de medida $\Delta\tau$ deve ser igual ou maior do que o tempo de percurso λ/2c. Assim, pondo $\Delta\tau \approx \lambda/2c$ em (1.11) obtemos,

$$d(\Delta L) = (c\hbar\lambda_{em}f/\pi \, \varepsilon \, P)^{1/2} \qquad (1.12).$$

De acordo com (1.4) concluímos que a intensidade mínima $h_{min}$ da OG que gera uma mudança de *comprimento efetivo* d(ΔL) deve obedecer a relação $h_{min} = d(\Delta L)/L = d(\Delta L)/(\lambda/2)$ de onde obtemos, usando (1.12):

$$h_{min} = 2(\hbar\lambda_{em}f^3/c\pi \, \varepsilon \, P)^{1/2} \qquad (1.13).$$

Supondo que P = 100 W, $\lambda_{em}$ = 500 nm, f = 1 kHz e ε = 0.3 obtemos, usando (1.13) a *sensibilidade* ou $h_{min} \approx 3 \cdot 10^{-22}$.

Como na estimativa vista acima de $h_{min}$ assumimos que a freqüência f da onda era de f = 1 kHz e que o percurso ótico L = λ/2 = c/2f = 150 km o raio de luz deverá efetuar um número n = 150/$L_b$ de passagens, onde $L_b$ é o comprimento do braço do interferômetro. O tempo $\Delta\tau$ de percurso do fóton no braço será igual a $\Delta\tau = \lambda/2c$ = 5 ms. No VIRGO como $L_b$ = 3 km teríamos n = 50. Nessas condições a variação de caminho ótico d(ΔL) = $h_{min}L$ = 4.5 $10^{-15}$ cm e a flutuação do número de fótons dN ≈ $N_o^{1/2}$, que segundo a (1.9), será dada por dN = $N_o^{1/2}$ = $\lambda_{em}/(2\pi \, d(\Delta L))$ ~$10^{10}$.

É importante notarmos que os interferômetros podem, em princípio, detectar OG com qualquer freqüência. A (1.13) mostra o valor de $h_{min}$ levando em conta, numa estimativa simples, somente alguns fatores tais como $\lambda_{em}$, potência P do laser, eficiência ε de detecção do sinal luminoso, etc...Entretanto, há muitos outros fatores que limitam a sensibilidade dos interferômetros e que dependem da freqüência f da OG. Estimativas e simulações muito precisas estão sendo feitas e já foram feitas de $h_{min}$ em função de f pelos observatórios GEO 600, LIGO, VIRGO, TAMA 300, AIGO, LISA e LISC levando em conta inúmeros efeitos. Estes são sísmicos, térmicos e termo−elásticos nos pêndulos que sustentam os espelhos, pressão da radiação, acústicos, eletromagnéticos, raios cósmicos, distorção das superfícies dos espelhos devido ao laser, revestimento dos espelhos, "shot noise",etc. Na Figura 2 mostramos o gráfico de $h_{min}(f)[1/sqrt(Hz)]$ em função de f [Hz] para o VIRGO onde os efeitos citados acima são levados em conta.

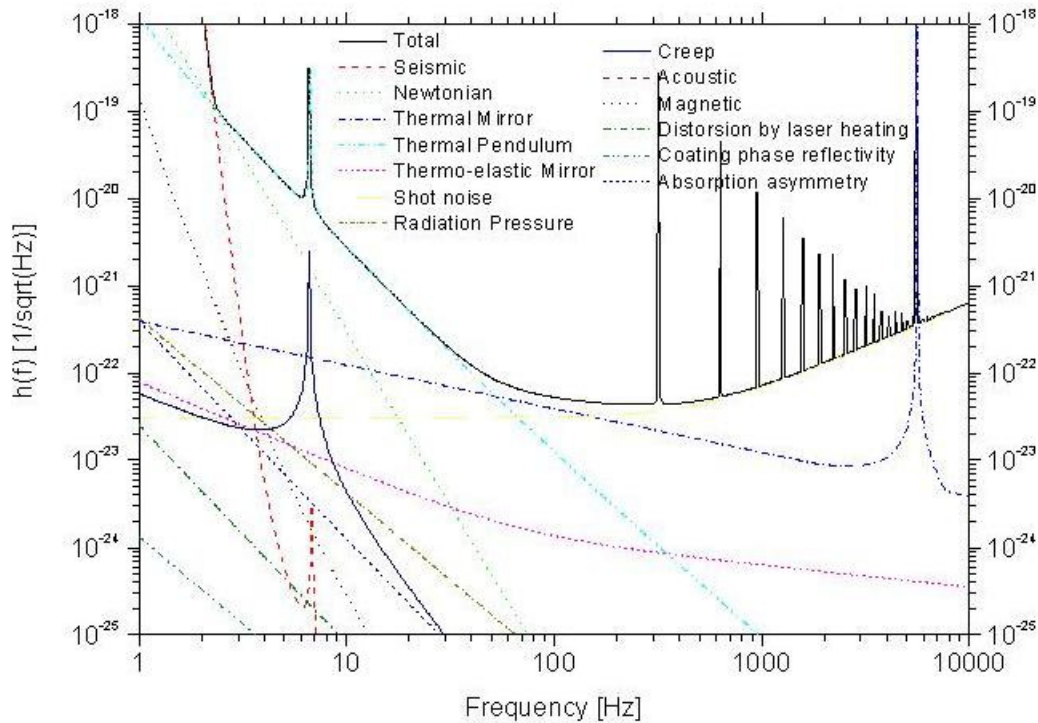

**Figura 2.** Curva da sensibilidade $h_{min}(f)[1/sqrt(Hz)]$ em função da freqüência f [Hz] prevista para o interferômetro VIRGO levando em conta limitações provocadas por muitos efeitos perturbativos.

Conforme artigo Giazotto et al.[11] a eficiência de detecção dos sinais da OG em VIRGO poderia ser muito ampliada aumentando o período de tempo T de observação da onda incidente. Para T ~3 anos, por exemplo, $h_{min}$~3 $10^{-22}$ estimado com (1.3) seria muito menor ~ $10^{-26}$.

## 2. Detectores Ressonantes.

Atualmente, os detectores ressonantes mais cotados para a detecção das OG são os constituídos por sólidos muito rígidos, em geral, em forma de barras cilíndricas ou esferas.[3,7,8] O primeiro detector desse tipo, inventado por Weber[12] em 1960, tinha o formato de um cilindro. Esses detectores sólidos que servem como antenas são barras cilíndricas ou esferas suspensas têm modos normais de vibração acústica com freqüências $\omega_n$ (n = 0,1,2,..). Quando uma OG incide sobre o sólido, com polarização e freqüência favoráveis, produz vibrações que podem ser medidas com o auxílio de transdutores conectados mecanicamente no referido sólido.[3,7,8,12,13] São usados transdutores e amplificadores de sinais eletrônicos que necessitam de cuidadoso projeto de tal modo a terem baixíssimos ruídos combinados com sistemas adequados de transferência de sinal. Amaldi et al.[13] usam um capacitor para detectar as deformações longitudinais na barra e um transformador supercondutor para acoplar os sinais induzidos a um d.c. SQUID[14,15] ("Superconducting Quantum Interference Device"). Existem atualmente vários laboratórios que estão usando ou fazendo projetos visando usar essa técnica ressonante: ALLEGRO (USA), AURIGA (Itália), EXPLORER (Itália), NIOBE (Austrália), miniGRAIL (Holanda) e GRAVITON (Brasil). Os quatro primeiros grupos usam barras cilíndricas e os dois últimos usam esferas. Informações sobre esses observatórios podem ser obtidas acessando o Google: "Resonant Gravitational Wave Detectors" ou "GEO 600".

Vejamos como estimar a sensibilidade de um detector ressonante cilíndrico. Poder−se−ia calcular as amplitudes de oscilações longitudinais na barra cilíndrica de uma maneira rigorosa levando em conta os seus modos normais estacionários[16] (vide Ohanian,[8] problema 12, pág.181). Vamos assumir que somente a freqüência fundamental $f_o = \omega_o/2\pi$ de uma onda estacionária longitudinal seja excitada pelas OG numa barra de comprimento $\ell$. Sendo $V_s$ a velocidade de propagação do som no material $f_o$ é dada[8,16] por $f_o = V_s/2\ell$. Como nas antenas usuais[7,8] $V_s \sim 6 \cdot 10^3$ m/s e $\ell \sim 3$ m vemos que $f_o \sim 1$ kHz. Entretanto, há uma maneira mais simples e que descreve razoavelmente bem os efeitos dessas oscilações da barra considerando−a como sendo um quadrupolo oscilante que estudamos antes.[5] Ele é formado por duas massas iguais M que oscilam ligados por uma mola com freqüência própria $\omega_o$ ao longo do eixo z, com amplitude a em torno de pontos de equilíbrio $\pm\ell$, ou seja, $z(t) = \pm \ell + a \sin(\omega_o t)$.

Conforme vimos no artigo anterior[6] uma OG *periódica*[5] que incide com amplitude h e freqüência angular $\omega_g = \omega$ polarizada ao longo do eixo z gera uma força F sobre uma massa M, ao longo de z, dada por $F = (M\omega^2 h\ell/2)\cos(\omega t)$. Nessas condições se o movimento harmônico for amortecido M obedecerá à equação:

$$M(d^2z/dt^2) + M\gamma(dz/dt) + m\omega_o^2(z-\ell) = (M\omega^2 h\ell/2)\exp(i\omega t) \qquad (2.1),$$

onde $\gamma$ é o fator de amortecimento associado às forças dissipativas agindo sobre o oscilador. No regime estacionário a amplitude deslocamento $\Delta\ell(t)$ gerada pela OG é dada por

$$\Delta\ell(t) = (\omega^2 h\ell/2)e^{i\omega t}/|\omega^2 - \omega_o^2 + i\gamma\omega| \qquad (2.2).$$

Como usualmente temos[7,8,13] $\omega_o \sim 2\pi\,10^3$ /s e $\gamma \sim 0.05$ s, ou seja, $\omega_o \gg \gamma$ a (2.2) pode ser aproximada por

$$\xi(t) = \Delta\ell(t) \approx (\omega_o h\ell/4)e^{i\omega t}/|\omega - \omega_o + i\gamma/2| \qquad (2.3)$$

De (2.3) verificamos que o valor máximo de $\Delta\ell(t)$ é obtido no caso ressonante, $\omega = \omega_o$, sendo dado então por $|\Delta\ell|_{max} = \omega_o h\ell/2\gamma = Qh\ell/2$, onde $Q = \omega_o/\gamma$ é o *"fator de qualidade"* do oscilador. Como para bons osciladores[3,7,8] $Q \sim 10^5$ vemos que $|\Delta\ell|_{max}$ seria muito maior do que o deslocamento $|\Delta\ell| = h\ell/2$ de uma partícula livre que é obtido de (2.2) fazendo $\omega_o = \gamma = 0$.

    Notemos que somente uma parte da energia incidente da OG é absorvida pela antena, a outra parte é espalhada, re-irradiada. Pode-se mostrar,[3,8] levando em conta as secções de choque de absorção e de espalhamento da OG, que a parte espalhada é desprezível (vide Apêndice). Desse modo assumiremos que a energia $E_g$ da OG é totalmente absorvida pela antena. O cálculo da sensibilidade da antena pode ser feito usando a seção de choque de absorção da energia incidente.[3,8] Entretanto, não seguiremos esse método. Adotaremos um procedimento mais simples que nos permite estimar aproximadamente a sensibilidade da barra ressonante.

    De acordo com (2.3) se a freqüência da onda for muito diferente de $\omega \approx \omega_o \pm \gamma/2$ o deslocamento $\Delta\ell$ cai drasticamente a zero. Assim, como temos $\omega_o \sim 2\pi\,10^3$ /s e $\gamma \sim 0.05$ s, somente quando $\omega_g = \omega \approx 6300 \pm 0.005$, ou seja, quando $\omega$ e $\omega_o$ concordarem em uma parte em $10^3$ teremos um $\Delta\ell$ não desprezível causada pela OG. Como essa tão precisa concordância de freqüências é muito improvável de acontecer (tendo em vista as emissões astrofísicas previstas[6]) há uma grande probabilidade de nenhuma OG ser detectada pela antena.

    Notemos que o deslocamento $\Delta\ell(t)$ dado por (2.3) foi obtido para uma OG *periódica*[5] num regime estacionário, ou seja, após a chegada da onda decorreu um intervalo de tempo $\Delta t > 1/\gamma \sim 20$ s. No regime estacionário a potência P da OG dada por $P = -dE_{OG}/dt$ é totalmente transformada em vibração de tal modo que $P = -dE_{OG}/dt = \gamma E_{vib}$, onde $E_{vib}$ é a energia de vibração acústica. Suponhamos agora que a OG seja emitida sob a forma de um *pulso* que tem uma distribuição de freqüências centrada

em $\omega_o$. Como espera–se que esses pulsos tenham uma duração[6] de tempo $\delta t$ ~$10^{-3}$ s o estado estacionário não será atingido pois $\Delta t > 1/\gamma$ ~ 20 s >> $\delta t$. Neste caso, então, o deslocamento $\Delta\ell$ gerado pelo pulso será, com certeza, muito menor do que o $\Delta\ell$ ressonante $|\Delta\ell|_{max} = Qh\ell/2$. Além disso, o sinal gerado sob forma de pulso poderá ser confundido com um ruído térmico.

Assim a detecção de uma OG *periódica* ou *pulsada* pela antena ressonante será muito difícil, pois $\Delta\ell$ gerado na antena será, provavelmente, muito pequeno. Como $\Delta\ell$ deve estar no intervalo $0 < \Delta\ell < Qh\ell/2$ vamos assumir[7] com *otimismo*, ou seja, na melhor das hipóteses, que $|\Delta\ell| \approx h\ell$. Portanto, teremos[5,7]

$$d\xi(t)/dt = d[\Delta\ell(t)]/dt \approx h\ell\omega e^{i\omega t} \qquad (2.4).$$

Nessas condições a energia vibracional $E_{vib}$ da antena que é igual a energia gravitacional incidente $E_g$ será dada por[7]

$$E_{vib} \sim M |d\xi(t)/dt|^2 = M(h\ell\omega_o)^2 = M h^2\ell^2\omega_o^2 \qquad (2.5),$$

onde M é a massa da barra ressonante. Suporemos que a barra esteja a uma temperatura T muito baixa T ~ 4 K de tal modo que a energia vibracional fundamental seja a única que esteja contribuindo para a energia térmica $E_t$ da barra dada por $E_t = kT$, onde k é a constante de Boltzmann. Para que a perturbação gerada pela OG seja detectável é preciso que $E_{vib}$ seja maior do que energia de agitação térmica $E_t$, isto é,

$$M h^2\ell^2\omega_o^2 > kT \qquad (2.6).$$

de onde verificamos que o valor mínimo de h, ou seja, $h_{min}$ deve obedecer à seguinte condição

$$h_{min} > (kT/M)^{1/2}/(\omega_o\ell) \qquad (2.7)$$

Assumindo que[7,8] M = 2.5 $10^3$ kg, $\omega_o = 2\pi 10^3$/s  e  $\ell$ = 2.5 m de (2.4) obtemos

$$h_{min} > 10^{-20} T^{1/2} \qquad (2.8).$$

Se a barra funcionar em temperaturas de mK a sensibilidade aumenta, podendo chegar até a $h_{min} > 10^{-21}$.

Existe uma limitação para a sensibilidade dos detectores ressonantes imposta pela Mecânica Quântica que é independente da temperatura: a energia gravitacional absorvida $E_g$ deve ser maior do que a energia de transição entre dois níveis quânticos $\hbar\omega_o$, isto é, a energia da barra deve mudar ao absorver energia dos "gravitons" pelo menos com a criação de um ou mais fônons. Assim, devemos ter $E_g > \hbar\omega_o$ que daria, usando (2.5),

$$h_{min} > (\hbar/M\omega_o)^{1/2}/\ell \qquad (2.9).$$

Assim colocando em (2.9) $M = 2.5 \cdot 10^3$ kg, $\omega_o = 2\pi \cdot 10^3$/s e $\ell = 2.5$ m, teremos $h_{min} > 10^{-20}$.

## 3. Conclusões.

Assumindo uma hipótese otimista (2.4), verificamos que a sensibilidade das antenas ressonantes é comparável com a dos interferômetros. A vantagem dos interferômetros é que eles poderiam detectar ondas com quaisquer freqüências. A antena ressonante tem várias desvantagens. Uma delas é que é projetada para detectar somente OG com uma determinada freqüência. Além disso, ela não é adequada para detectar pulsos de OG com duração $\delta t \ll 1/\gamma \sim 20$ s, pois os sinais gerados poderiam ser confundidos com ruídos térmicos. Essa última dificuldade, entretanto, pode ser contornada acoplando pelo menos duas antenas em coincidência. As potencialidades técnicas dos interferômetros são imensamente superiores às usadas nos sólidos ressonantes. Acreditamos que se as OG existirem elas só poderão ser detectadas *com certeza* pelos interferômetros. A detecção de OG pelos sistemas ressonantes será fortuita.

**APÊNDICE. Secção de Choque de Absorção e Espalhamento de OG.**

Para calcular a sensibilidade de um sólido ressonante pode−se usar um parâmetro denominado de *secção de choque*.[3,8] Vejamos primeiramente a *seção de choque de espalhamento* ou de *"scattering"*. Como as OG excitam vibrações quadrupolares no detector essas oscilações por sua vez irradiam OG; isto é, uma parte da energia é absorvida e a outra é re−irradiada em todo espaço como uma nova onda. A secção de choque de "scattering" $\sigma_{scatt}$ é definida como a razão da potência re−irradiada pelo fluxo de energia incidente $F = \Phi$[5]:

$\sigma_{scatt}$ = < potência re−irradiada >/< fluxo de energia incidente >   (A.1),

onde os colchetes <...> indicam uma média num ciclo. A $\sigma_{scatt}$, que tem dimensão de área, mede a eficiência com que a antena espalha em todas as direções a onda incidente.

O fluxo médio de energia incidente < F > com uma única polarização (+), portanto só com $h_+ = h$, é dado pela (I.3) do artigo anterior[6]

$$< F > = (c^3/32\pi G) h^2 \omega^2 \qquad (A.2)$$

Como a potência re−irradiada $< dE/dt >_{rad}$ é obtida integrando a $(E.5)^6$ em todo o espaço e efetuando uma média num ciclo temos,

$$< dE/dt >_{rad} = - (16G/15c^5) (M\ell\Delta\ell)^2 \omega^6 \qquad (A.3),$$

lembrando que $\ell$ é o comprimento da antena e $\Delta\ell$ a amplitude da deformação gerada pela OG na antena. De acordo com (A.1) a $\sigma_{scatt}$ é dada por $\sigma_{scatt} = < dE/dt >_{rad}/< F >$. Levando em conta que $\Delta\ell$ é definida por (2.3) e usando (A.1)−(A.3) obtemos

$$\sigma_{scatt} = (128\pi G^2/15c^8)(M\ell^2\omega_o^2)^2/[(\omega-\omega_o)^2 + (\gamma/2)^2] \qquad (A.4).$$

A secção de choque de absorção de energia da OG $\sigma_{abs}$ é definida por

$$\sigma_{abs} = < dE/dt >_{abs}/ < F > \qquad (A.5),$$

onde a potência absorvida $P_{abs} = < dE/dt >_{abs}$ é dada, conforme vimos na Seção 2, por $P_{abs} = dE_{OG}/dt = \gamma E_{vibr}$. Assim, escrevendo (A.5) na forma $\sigma_{abs} = \gamma E_{vibr} / < F >$, lembrando que $E_{vib} = 2 M(d\xi/dt)$ onde $\xi(t)$ é definida pela (2.3) obtemos

$$\sigma_{abs} = (8\pi/15) (M\gamma\ell^2\omega_o^2)/ [(\omega-\omega_o)^2 + (\gamma/2)^2] \qquad (A.6).$$

De (A.4) e (A.6) temos $\sigma_{abs} = (\gamma/\gamma_{rad})\sigma_{scatt}$, onde $\gamma_{rad} = (16G/15c^5)M\ell^2\omega_o^4$. Como $G = 6.67\ 10^{-8}$ cgs, $c = 3\ 10^{10}$ m/s, $\gamma \sim 0.05$ /s, $M \sim 2\ 10^3$ kg e $\omega_o \sim 2\pi\ 10^3$ /s vemos que $\sigma_{scatt} \sim 10^{-35}\sigma_{abs}$ mostrando que praticamente não há espalhamento da OG pela antena, conforme dissemos na Seção 2.